\documentclass[12pt]{article}
\usepackage{color}

\newcommand{\beq}{\begin{equation}}
\newcommand{\eeq}{\end{equation}}
\newcommand{\beqa}{\begin{eqnarray}}
\newcommand{\eeqa}{\end{eqnarray}}
\newcommand{\beqar}{\begin{eqnarray*}}
\newcommand{\eeqar}{\end{eqnarray*}}

\newcommand{\al}{\alpha}
\newcommand{\be}{\beta}

\def\non          {\nonumber}

\def\Tr           {\mbox{\rm Tr}\,}

\def\cd           {{\cdot}}
\def\ran          {\rangle}
\def\lan          {\langle}
\def\fsk    {k\!\!\!\!/\,}
\def\fsH    {H\!\!\!\!/\,}

\newcommand{\ga}{\gamma}

\newcommand{\lam}{\lambda}

\newcommand\bPsi{{\bar \Psi }}

\newcommand{\z}{\zeta}

\newcommand{\labell}[1]{\label{#1}} 
\newcommand{\reef}[1]{(\ref{#1})}

\newcommand\veps{\varepsilon}

\newcommand\cD{{\cal D}}

\newcommand\bu{\bar{u}}

\def\sst#1{{\scriptscriptstyle #1}}
\def\0{{\sst{(0)}}}
\def\1{{\sst{(1)}}}
\def\2{{\sst{(2)}}}
\def\3{{\sst{(3)}}}
\def\4{{\sst{(4)}}}
\def\5{{\sst{(5)}}}
\def\6{{\sst{(6)}}}
\def\7{{\sst{(7)}}}
\def\8{{\sst{(8)}}}

\begin{document}
\baselineskip 18pt%
\begin{titlepage}
\vspace*{1mm}%
\hfill
\vbox{

    \halign{#\hfil         \cr
           } 
      }  
\vspace*{7mm}

\center{ {\bf \Large  SuperYang-Mills, Chern-Simons couplings and their all order $\alpha'$
corrections in IIB  superstring theory
}}\vspace*{1mm} \centerline{{\Large {\bf  }}}
\begin{center}
{Ehsan Hatefi }$\footnote{E-mail:ehatefi@ictp.it.
}$

\vspace*{0.6cm}{ {\it
International Centre for Theoretical Physics, Strada Costiera 11, Trieste, Italy,
\\ and\\
Simons Center for Geometry and Physics, Stony Brook University,\\
Stony Brook, NY 11794, USA
}}
\vspace*{0.1cm}
\vspace*{.1cm}
\end{center}
\begin{center}{\bf Abstract}\end{center}
\begin{quote}

We explore the closed  form of the correlation function of four spin operators (including one closed string Ramond-Ramond (RR) and two open string fermions) and one current in ten dimensions, to be able to find the complete  and the closed form of the amplitude of one closed string Ramond-Ramond, one gauge field and  two fermionic strings (with the same chirality)  to all orders in $\alpha'$ in IIB superstring theory. In particular  we use  a special gauge fixing to the amplitude  and apply fermions' equations of motion  to  $<V_{C} V_{A}V_{\bar\psi}V_{\psi} >$ correlator. String amplitude induced that neither there should  be any $u-$channel gauge poles for $p=n+2$ case  nor there are couplings between two fermions and two gauge fields for $p=n$ case  in the field theory of type IIB.

All infinite $u-$ channel scalar poles and $t,s-$channel fermion poles of the string amplitude are useful in discovering new couplings of type IIB. More specifically, by making use of the  SYM couplings of one scalar, one gauge and two fermions and their all order $\alpha'$ higher derivative corrections, we are able to exactly produce all infinite  $(s+t+u)-$ channel scalar poles of $<V_{C} V_{A}V_{\bar\psi}V_{\psi} >$ .

\end{quote}
\end{titlepage}

\section{Introduction}

D-branes  \cite{Polchinski:1995mt},\cite{Witten:1995im},\cite{Polchinski:1996na}
are fundamental non perturbative objects in string theory. They play the key role in diverse subjects  as well as in superstring theories. To be able to talk about the derivation of  Ads/CFT one has to deal with these fundamental objects. Let us point out one of their dynamical aspects. In order to be able to describe the different transitions of open/closed strings, we consider an interesting paper\cite{Ademollo:1974fc}.
 \vskip.1in

 For the completeness, to get familiar with string dualities and to observe various dual descriptions, we introduce \cite{Polchinski:1996nb} to the interested reader. As an example one may talk about a particular configuration such as $D0/D4$ system where its importance as well as its applications/explanations are addressed  in  \cite{Hatefi:2012sy}.
  \vskip.1in
 To be more specific, one can see the presence of  world volume theory from supergravity point of view \cite{Hatefi:2012bp}. Basically in  \cite{Hatefi:2012bp}, we have just employed a new version of ADM reduction and it is applied to type IIB superstring theory. Indeed this kind of ADM  reduction has to be deduced to five dimensional hyperboloidal space. In this particular formalism we could understand the appearance of either an Ads or ds space.

 \vskip .1in

 One has to apply suitable  boundary conditions \cite{Polchinski:1994fq}, to be able to observe that D$_p$-branes must be seen as some hypersurfaces in decompactification space ( ten dimensions of flat space time) where $p$ is interpreted as spatial dimension of a  D$_p$-brane. IIA (IIB) does include BPS D$_p$-branes with even (odd)  $p$ and indeed supersymmetry is guaranteed. BPS branes also  carry RR(C-field) charge. What can we say about the dynamical aspects of branes? In order to deal with dynamical aspects of D$_p$-branes  we need to discover the general form of the  effective actions. Basically  we might work with either  bosonic effective actions or their supersymmetric  versions and these bosonic effective actions for diverse brane configurations have already been found in \cite{Myers:1999ps}.

\vskip 0.2in

One should argue that  the supersymmetrized  versions of the bosonic actions \cite{Myers:1999ps} have not been entirely explored yet, nevertheless we point out to an interesting and pioneering work \cite{Howe:2006rv}. Let us just highlight the most important references. In order to talk about the effective action of just a single bosonic   D$_p$-brane \cite{Leigh:1989jq} must be taken into account. To work with the supersymmetric action of a  D$_p$-brane, we refer to  \cite{Cederwall:1996pv}.
 \vskip .2in

It is worth mentioning that  Myers terms, the Chern-Simons, Wess-Zumino (WZ) and Born-Infeld action are completely derived in \cite{Tseytlin:1999dj,Tseytlin:1997csa,Hatefi:2010ik,Hatefi:2012zh}. More importantly, we have explained how to look for all the standard effective field theory methods of Myers terms, Taylor and Pull-back where their all order $\alpha'$ corrections are found out in \cite{Hatefi:2012wj}. However, we discussed in \cite{Hatefi:2012zh} that certainly the presence of some other methods apart from those three standard ways is needed. In fact in order to obtain all the infinite higher derivative corrections in string theory one should go further and construct new methods for both BPS
\cite{Hatefi:2012zh,Hatefi:2013eia}  and non-BPS branes \cite{Hatefi:2013yxa,Hatefi:2013mwa,Hatefi:2012cp,Garousi:2007fk,Hatefi:2008ab}.

\vskip 0.2in

 Open strings have provided various features  and their importance can be extracted by dealing with  some formal aspects of scattering amplitude arguments . For instance we point out to two different conjectures on taking  quantum effects of the BPS strings \cite{Park:2007mc} and a given prescription of all order $\alpha'$ higher derivative corrections to BPS and non-BPS branes \cite{Hatefi:2012rx} for which  the first conjecture shows the effects on the  curvature of the host branes.
\vskip 0.2in

We have been following several motivations for our continuous works \cite{Hatefi:2012zh}, so let us emphasize some of them
once more. Several important results appeared in \cite{Koerber:2002zb,Keurentjes:2004tu,Denef:2000rj} and  we want to explore the entire form of the effective actions to describe dynamical aspects of branes.  One of the goals of this paper is to actually prepare more information on the general structure of all order Chern-Simons and Born-Infeld effective actions in superstring theory.
For the completeness we highlight  \cite{Hashimoto:1996bf} and consider some of the papers which are related  to either  scattering of BPS branes or related to the formal applications of the BPS and non-BPS branes \cite{Hashimoto:1996kf,Lambert:2003zr,Dudas:2001wd,Antoniadis:1999xk,deAlwis:2013gka}.

\vskip 0.1in

It is important to have some tools to be able to work with  higher point functions of the superstring theory
, given the fact that the derivation of  AdS/CFT correspondence is unknown. Because we know that by going through them and inside the  Ads/CFT  there exists a very straightforward  relation  of a closed and an open string. So  by working with mixture open-closed  amplitudes one might hope to shed light in understandings  all order $\alpha'$ corrections and to the other future works. For instance, we have recently derived various recent  WZ effective couplings  with their all order $\alpha'$ higher derivative corrections  and just showed  that those corrections  must be taken into account to clarify  the $N^3$ entropy of M5 brane, where for further explanations we just refer   \cite{Hatefi:2012sy}  to the reader.

\vskip 0.1in

In fact  $\alpha'$ corrections  coming from  all the infinite couplings of the branes with lower dimensions with RR field. Therefore one understands the dissolution of
soliton objects or lower dimensional branes \cite{Hatefi:2012sy} within branes with higher dimensions. One might talk about a particular application to Myers terms as follows.
In the system of $D(-1)/D3$ , one  employs  higher order $\alpha'$ Myers terms to be able to explore
this configuration carries  $N^2$ entropy relation. To observe several applications to some of the recent $\alpha'$ corrections,  to WZ couplings in  flux vacua and M-theory  references \cite{Maxfield:2013wka,
McOrist:2012yc,Hatefi:2012wz} are worth considering.

\vskip 0.2in

If we  try to work out some of the recent works on  Myers terms and new kind of WZ effective actions \cite{Hatefi:2010ik,Hatefi:2012cp,Garousi:2007fk,Hatefi:2008ab,Hatefi:2012rx,Ferrari:2013pi,Hatefi:2012ve} then we will be able to find out  some of the corrections in string theory \cite{Boels:2010bv}.  Basically in order to produce all infinite scalar poles of the amplitude of  $<V_{C} V_{A}V_{\bar\psi}V_{\psi} >$, an infinite number of $\alpha'$ corrections to  two fermion-one scalar one gauge field couplings is needed. The other important point is as follows.
Unlike $<V_{C} V_{\phi}V_{\bar\psi}V_{\psi} >$ correlators , the closed form of the  $<V_{C} V_{A}V_{\bar\psi}V_{\psi} >$   amplitude  includes just infinite $u-$channel scalar poles in its final form and the direct computations of this paper shows that there is no even one single $u-$ channel gauge field pole left over.

\vskip 0.1in

Direct computations of this paper shows that there are no corrections to two fermions and two gauge field couplings of type IIB for $p=n$ case where $n$ is the rank of RR field strength $(H)$. This fact is in favor of carrying out direct conformal field theory techniques , instead of doing T-duality transformation to  $<V_{C} V_{\phi}V_{\bar\psi}V_{\psi} >$ amplitude,  see \cite{Hatefi:2012zh} for more promising reasons  \footnote{
 As it is seen the entire result of  $<V_{C}V_{\phi} V_AV_A>$ can not be derived by applying T-duality transformation to $<V_{C}V_A V_AV_A>$ of \cite{Hatefi:2010ik}. } .

\vskip 0.1in

If we find the  infinite  corrections to one scalar,one gauge and two fermion fields , then we are able to find all infinite  scalar $(t+s+u)$-channel poles of the string theory  for   $p+2=n$ case. Hence by comparing all infinite scalar poles of the string theory amplitude in  $(t+s+u)$-channel with   field theory vertices we obtain  an infinite corrections of  two fermions,one on-shell gauge and one off-shell scalar field . It is important to highlight the fact that these all order corrections of type IIB can not be used in type IIA and may have diverse application to either M-theory
 \cite{Hatefi:2012sy,McOrist:2012yc,Hatefi:2012wz} or  F-theory  \cite{Vafa:1996xn}.

\vskip 0.2in

 Therefore  this paper among the other things clearly indicates that SYM vertex operators including one on-shell gauge,one off-shell scalar field and two fermion fields will give rise exactly all the same infinite scalar poles that appeared in string amplitude of  $<V_{C} V_{A} V_{\bar\psi}V_{\psi} >$ as well.

\vskip 0.3in

Here is the outline of the paper.

 In the second section we carry out direct conformal field theory techniques to be able to find out the entire form of the amplitude of  a closed string RR ,one gauge field and  two fermion fields  $<V_{C} V_{A} V_{\bar\psi}V_{\psi} >$. Note that in this paper both fermions carry the same chirality , hence the calculations of this paper work just for  IIB  and the corrections that we are getting to derive in IIB theory can not be applied to IIA.

\vskip 0.1in

 We also expand our S-matrix to be able to derive all infinite  extensions to the unknown vertices.  In particular  we explicitly show that there are just infinite scalar  $u$-channel massless poles for  $p+2=n$ case. Indeed all infinite gauge poles that were appeared in $<V_{C} V_{\phi} V_{\bar\psi}V_{\psi} >$ for $p=n$ case, are disappeared in the closed form of  $<V_{C} V_{A} V_{\bar\psi}V_{\psi} >$ and this is an interesting fact in favor of carrying out  direct computations.  Then we  try to produce all infinite scalar poles of the amplitude in $(t+s+u)$ channel poles. We also  explicitly write down all the desired couplings of two fermions and two gauge fields and show that they do not match with string theory amplitude . This clearly confirms that there are no corrections to two fermion-two gauge field couplings of type IIB. We also produce  all infinite  $t,s$-channel fermion poles involving their extensions to new couplings.

Eventually  we  conclude and point out fascinating relation between open -closed string amplitudes. One might look at  Appendix A and B of \cite{Hatefi:2012wj,Hatefi:2012ve} to get used to standard notations and some other details.
\vskip 0.2in

It is worth mentioning that,  this paper may shed light in understanding the  universal behavior of all order $\alpha'$ higher derivative corrections of the string theory \cite{Hatefi:2012rx}.  Having used the results of this paper, we are able to judge that  a universal conjecture  given in \cite{Hatefi:2012rx} does apply even to fermionic S-matrices. This conjecture may also be applied  to obtain all the infinite singularities of five and six point BPS functions without any knowledge of world sheet integrals.

\vskip 0.3in

\section{  Complete form of  $RR A  \bar\psi \psi$ amplitude in  type IIB}

To obtain the entire and closed form of the S-matrix elements of one closed string RR (C-field), one gauge field and two fermions with the same chirality ( which makes sense in the world volume of type IIB ), one has to deal with the direct CFT techniques. It is worth addressing several important references on superstring theory \cite{Witten:2013tpa,Witten:2013cia,
D'Hoker:2013eea}, higher point BPS functions
\cite{Hatefi:2010ik,Hatefi:2012rx,Hatefi:2012ve,Bilal:2001hb,Barreiro:2012aw,Barreiro:2013dpa,Stieberger:2009hq,Kennedy:1999nn,Chandia:2003sh} and non-BPS \cite{Hatefi:2012wj,Hatefi:2013mwa} tree level calculations.

For the completeness we point out the relevant vertex operators for our computations as follows
\beqa
V_{A}^{(0)}(x) &=& \xi_{a}\bigg(\partial
X^a(x)+\alpha' ik\cd\psi\psi^a(x)\bigg)e^{\alpha' ik\cd X(x)},
\nonumber\\
V_{A}^{(-2)}(y) &=&e^{-2\phi(y)}V_{A}^{(0)}(y),
\nonumber\\
V_{\bPsi}^{(-1/2)}(x)&=&\bu^Ae^{-\phi(x)/2}S_A(x)\,e^{\alpha'iq.X(x)} \nonumber\\
V_{\Psi}^{(-1/2)}(x)&=&u^Be^{-\phi(x)/2}S_B(x)\,e^{\alpha'iq.X(x)} \nonumber\\
V_{C}^{(-\frac{1}{2},-\frac{1}{2})}(z,\bar{z})&=&(P_{-}\fsH_{(n)}M_p)^{\al\be}e^{-\phi(z)/2}
S_{\al}(z)e^{i\frac{\alpha'}{2}p\cd X(z)}e^{-\phi(\bar{z})/2} S_{\be}(\bar{z})
e^{i\frac{\alpha'}{2}p\cd D \cd X(\bar{z})},
\label{d4Vs}
\eeqa
Majorana-Weyl wave function $u^A$ is  also introduced in ten dimensions.
The on-shell condition  for RR, fermion and scalar field is
$p^2=q^2=k^2=0 $. For the other notations on charge conjugation, the definition of the traces and field strength of RR in IIB, reference \cite{Hatefi:2013eia} should be considered. Note also that in order to work with holomorphic propagators,
  doubling tricks were used  \cite{Hatefi:2012wj}, however, let us point out the standard correlators

\begin{eqnarray}
\lan X^{\mu}(z)X^{\nu}(w)\ran & = & -\frac{\alpha'}{2}\eta^{\mu\nu}\log(z-w) , \non \\
\lan \psi^{\mu}(z)\psi^{\nu}(w) \ran & = & -\frac{\alpha'}{2}\eta^{\mu\nu}(z-w)^{-1} \ ,\non \\
\lan\phi(z)\phi(w)\ran & = & -\log(z-w) \ .
\labell{prop}\end{eqnarray}

  One gauge field and two fermions  amplitude $< V_{A} V_{\bar\psi}V_{\psi} >$
is computed  in \cite{Polchinski:1998aa}. If we normalize its S-Matrix  with   $( iT_p 2^{1/2}\pi\alpha')$ coefficient, then
one can show that the S-matrix should be produced by extracting the kinetic term of the fermion fields  and taking into account the commutator
in $D^a\psi$ as follows
\beqa
 (2\pi\alpha' T_p)\Tr(\bar\psi\ga^{a} D_{a}\psi), \quad D^a\psi=\partial^a\psi-i[A^a,\psi]\nonumber\eeqa

\vskip.1in

The closed form of our amplitude  is given by taking the closed form of the following  correlator

 \beqa
  <  V_{A}^{(0)}{(x_{1})}
V_{\bar\psi}^{(-1/2)}{(x_{2})}V_{\psi}^{(-1/2)}{(x_{3})}
V_{RR}^{(-\frac{1}{2},-\frac{1}{2})}(z,\bar{z}) >  \label{sstring}\eeqa

 Notice that the complete result of our calculations should not be used for IIA as here we are considering both fermions with the same chirality, thus all order corrections in this paper can not be used for IIA.

\vskip.1in

The amplitude has two different parts, for the first part we need to have the correlation function of four spin operators in ten dimensions  \cite{Friedan:1985ge,Hartl:2010ks}  as

\beqa
  I_{1\gamma\delta\alpha\beta}&=&<S_{\gamma}(x_2)S_{\delta}(x_3)S_{\alpha}(x_4)S_{\beta}(x_5)>=
 \bigg[(\gamma^\mu C)_{\alpha\beta}(\gamma_\mu C)_{\gamma\delta} x_{25}x_{34}
              -(\gamma^\mu C)_{\gamma\beta}(\ga_\mu C)_{\alpha\delta} x_{23} x_{45} \bigg]\nonumber\\&&\times
                \frac{1}{2 (x_{23} x_{24} x_{25} x_{34} x_{35} x_{45})^{3/4}}              \nonumber\eeqa
with $x_{ij}=x_i-x_j, x_4=z=x+iy, x_5=\bar z=x-iy$.
Having replaced   $ I_{1\gamma\delta\alpha\beta}$ in the first part of the amplitude, we obtain

\beqa {\cal A}_{1}^{C A \bar\psi \psi}& \sim &
 \int
 dx_{1}dx_{2}dx_{3}dx_{4} dx_{5}\,
(P_{-}\fsH_{(n)}M_p)^{\alpha\beta}\xi_{1a} \bu_1^{\gamma} u_2^{\delta}  (x_{23}x_{24}x_{25}x_{34}x_{35}x_{45})^{-1/4}
\nonumber\\&&\times I_{1\gamma\delta\alpha\beta} I_2 I_3^a
 \Tr(\lam_1\lam_2\lam_3),\labell{125}\eeqa

with
\beqa
I_2 &=&|x_{12}|^{\alpha'^2 k_1.k_2}|x_{13}|^{\alpha'^2 k_1.k_3}|x_{14}x_{15}|^{\frac{\alpha'^2}{2} k_1.p}|x_{23}|^{\alpha'^2 k_2.k_3}|
x_{24}x_{25}|^{\frac{\alpha'^2}{2} k_2.p}
|x_{34}x_{35}|^{\frac{\alpha'^2}{2} k_3.p}|x_{45}|^{\frac{\alpha'^2}{4}p.D.p}\nonumber\eeqa
\beqa
 I_3^{a}&=& ik_2^a \bigg(\frac{x_{42}}{x_{14}x_{12}}+ \frac{x_{52}}{x_{15}x_{12}}\bigg)+ik_3^a \bigg(\frac{x_{43}}{x_{14}x_{13}}+ \frac{x_{53}}{x_{15}x_{13}}\bigg)
\nonumber\eeqa

\vskip 0.1in

Now  we can obviously see that the amplitude  has SL(2,R) invariance property. We shall use a very specific gauge fixing as
  $(x_1=0,x_2=1,x_3=\infty)$ and in particular we are going to employ the following definitions for  Mandelstam variables

\beqa
s&=&-\frac{\alpha'}{2}(k_1+k_3)^2, \quad t=-\frac{\alpha'}{2}(k_1+k_2)^2, \quad u=- \frac{\alpha'}{2}(k_3+k_2)^2
\nonumber
\eeqa
 Thus having gauge fixed  it, one can find  the complete form of the first part of the amplitude  as follows

\beqa {\cal A}_{1}^{C A\bar\psi \psi}& \sim & (P_{-}\fsH_{(n)}M_p)^{\alpha\beta}\xi_{1a} \bu_1^\gamma u_2^\delta (\frac{-1}{2}) \int\int
 dz d\bar z |z|^{2t+2s}|1-z|^{2t+2u-2} (z-\bar z)^{-2(t+s+u)-1},
\nonumber\\&&\times  \bigg[(\ga^\mu C)_{\ga\delta}(\ga_\mu C)_{\alpha\beta}(1-\bar z)+(z-\bar z)(\ga^\mu C)_{\ga \beta}(\ga_\mu C)_{\alpha \delta}\bigg] \nonumber\\&&\times \bigg(2ik_2^a-\frac{(z+\bar z)(ik_2^a+ik_3^a)}{|z|^2}\bigg)\Tr(\lam_1\lam_2\lam_3)
\labell{amp3q},\eeqa

\vskip 0.1in

 In order to find out the complete result of the amplitude to all orders in $\alpha'$ one has to take integrations on the location of closed string RR where all details on these integrals can be partially seen in \cite{Fotopoulos:2001pt}  and they can be  completely found in the  Appendix B of \cite{Hatefi:2012wj}. Therefore the complete form of the first part of the amplitude is appeared as

\beqa
{\cal A}_{1}^{C A\bar\psi \psi}& \sim& (P_{-}\fsH_{(n)}M_p)^{\alpha\beta}\xi_{1a} \bu_1^\gamma u_2^\delta (\frac{-1}{2})
  \bigg\{ (\gamma^\mu C)_{\gamma\delta}(\gamma_\mu C)_{\alpha\beta} \bigg[ik_2^a( us L_1-2s L_2)
  -ik_3^a  ( ut L_1-2t L_2)\bigg]\nonumber\\&&+(\gamma^\mu C)_{\gamma \beta}(\gamma_\mu C)_{\alpha \delta} (2ik_2^a us -2ik_3^a ut) L_1\bigg\} \Tr(\lam_1\lam_2\lam_3)
\labell{amp3q},\eeqa

with
\beqa
L_1&=&(2)^{-2(t+s+u)}\pi{\frac{\Gamma(-u)
\Gamma(-s)\Gamma(-t)\Gamma(-t-s-u+\frac{1}{2})}
{\Gamma(-u-t+1)\Gamma(-t-s+1)\Gamma(-s-u+1)}},\nonumber\\
L_2&=&(2)^{-2(t+s+u)-1}\pi{\frac{\Gamma(-u+\frac{1}{2})
\Gamma(-s+\frac{1}{2})\Gamma(-t+\frac{1}{2})\Gamma(-t-s-u)}
{\Gamma(-u-t+1)\Gamma(-t-s+1)\Gamma(-s-u+1)}}
\label{Ls}
\eeqa

\vskip.2in

 Unlike the first part of $C \phi\bar\psi \psi$ amplitude , this  part does not seem to have any $u$- channel gauge nor scalar poles.
 Let us talk about the second part of the $C A\bar\psi \psi$ amplitude. For this part one needs to have the closed form of the correlation function
 of four spin operators (with the same chirality) and one current. The details for deriving this correlator have been given in section 2. of \cite{Hatefi:2013eia}, however, for the completeness here we are going to write down the complete form of that correlator as follows:

\beqa
< :\psi^b \psi ^a(x_1) :S_{\alpha}(x_2):  S_{\beta}(x_3): S_{\gamma}(x_4):  S_{\delta}(x_5) :> &=& I^{ba}_{\alpha\beta\gamma\delta}\nonumber\eeqa

with

\beqa
I^{ba}_{\alpha\beta\gamma\delta}&=&\frac{(x_{23}  x_{24}  x_{25}  x_{34}  x_{35}  x_{45})^{-3/4}}{4 (x_{12}  x_{13}  x_{14}  x_{15})}\bigg[  \bigg((\gamma^b  C)_{\gamma \beta}  (\gamma^a C)_{\alpha \delta} -(\gamma^b  C)_{\alpha \delta}  (\gamma^a  C)_{\gamma \beta}\bigg)  (  x_{12}  x_{14}  x_{35}  - x_{15}  x_{13}  x_{24})
\nonumber\\&&\times x_{23}  x_{45}
-   \bigg( (\gamma^b  C)_{\alpha \beta}  (\gamma^a C)_{\gamma \delta} + (\gamma^b  C)_{\gamma \delta}  (\gamma^a  C)_{\alpha \beta}\bigg) x_{25}  x_{34} (  x_{15}  x_{13}  x_{24} + x_{14}  x_{12}  x_{35} ) \nonumber\\&&
 +   (\Gamma^{ba \lambda}  C)_{\alpha \beta}  (\gamma_\lambda  C)_{\gamma \delta}  x_{23}  x_{25}  x_{34} (  x_{14}  x_{15} )
+(\Gamma^{ba \lambda}  C)_{\gamma \delta}  (\gamma_\lambda  C)_{\alpha \beta}  x_{25} x_{34}  x_{45} (  x_{12}  x_{13} ) \nonumber\\&&
-  (\Gamma^{ba \lambda}  C)_{\alpha \delta}  (\gamma_\lambda  C)_{\gamma \beta} x_{23}   x_{25}   x_{45} (  x_{14}   x_{13})
+  (\Gamma^{ba \lambda}  C)_{\gamma \beta}   (\gamma_\lambda  C)_{\alpha \delta}  x_{23}   x_{34}   x_{45}
 (  x_{12}  x_{15} )\bigg] \label{esi1}
\eeqa

\vskip 0.1in

By substituting the second part of the  gauge field 's vertex operator (in zero picture) and taking above correlator into the amplitude we find

 \beqa {\cal A}_{2}^{C A\bar\psi \psi} \sim  \int
 dx_{1}dx_{2}dx_{3}dx_{4} dx_{5}\,
(P_{-}\fsH_{(n)}M_p)^{\gamma\delta}\xi_{1a} (2ik_{1b})\bu_1^{\alpha} u_2^{\beta} ( x_{23}x_{24}x_{25}x_{34}x_{35} x_{45})^{-1/4}I_2  I^{ba}_{\alpha\beta\gamma\delta}\labell{1289}\eeqa
\vskip.1in

Concerning the above gauge fixing  and evaluating  the integrals on closed string position, the closed form of the second part
of our S-matrix is given by

\beqa {\cal A}_{2}^{C A \bar\psi \psi}  &\sim &  \bigg({\cal A}_{21}+{\cal A}_{22}+{\cal A}_{23}+{\cal A}_{24}+{\cal A}_{25}+{\cal A}_{26}\bigg)\Tr(\lam_1\lam_2\lam_3)\nonumber\eeqa

meanwhile
\beqa
{\cal A}_{21} &=&\frac{1}{4}(P_{-}\fsH_{(n)}M_p)^{\gamma\delta}\xi_{1a}(2ik_{1b}) \bu_1^{\alpha} u_2^{\beta} (\Gamma^{ba\lambda } C)_{
\alpha\beta}(\gamma_\lambda C)_{\gamma\delta}
\bigg[L_2+\frac{u}{2}(t+s)L_1\bigg]
\nonumber\\
\nonumber\\
{\cal A}_{22} &=&\frac{-1}{4}(P_{-}\fsH_{(n)}M_p)^{\gamma\delta}\xi_{1a}(2ik_{1b}) \bu_1^{\alpha} u_2^{\beta} (\Gamma^{ba\lambda } C)
_{\gamma\delta}(\gamma_\lambda C)_{\alpha\beta}
\bigg[L_1(st)+\frac{1}{2}L_3\bigg]
\nonumber\\
{\cal A}_{23} &=&\frac{-1}{4}(P_{-}\fsH_{(n)}M_p)^{\gamma\delta}\xi_{1a}(2ik_{1b}) \bu_1^{\alpha} u_2^{\beta} (\Gamma^{ba\lambda } C)
_{\alpha\delta}(\gamma_\lambda C)_{\gamma\beta}
\bigg[-L_1(su)+\frac{1}{2}L_3\bigg]
\nonumber\\
{\cal A}_{24} &=&\frac{1}{4}(P_{-}\fsH_{(n)}M_p)^{\gamma\delta}\xi_{1a}(2ik_{1b}) \bu_1^{\alpha} u_2^{\beta} (\Gamma^{ba\lambda } C)
_{\gamma\beta}(\ga_\lambda C)_{\alpha\delta}
\bigg[L_1(tu)-\frac{1}{2}L_3\bigg]
\nonumber\\
{\cal A}_{25} &=&\frac{1}{4}(P_{-}\fsH_{(n)}M_p)^{\gamma\delta}\xi_{1a}(2ik_{1b}) \bu_1^{\alpha} u_2^{\beta}
\bigg[u(s+t)L_1+L_3\bigg]
\nonumber\\&&\times\bigg((\ga^{b} C)_{\ga\beta}(\ga^a C)_{\alpha\delta}-(\ga^{b} C)_{\alpha\delta}(\ga^a C)_{\ga\beta}\bigg)
\nonumber\\
\nonumber\\
{\cal A}_{26} &=&\frac{-1}{4}(P_{-}\fsH_{(n)}M_p)^{\ga\delta}\xi_{1a}(2ik_{1b}) \bu_1^{\alpha} u_2^{\beta}
\bigg[-\frac{1}{2}u(s-t)L_1+(-t+s)L_2
\bigg]
\nonumber\\&&\times\bigg(-(\ga^{a} C)_{\ga\delta}(\ga^b C)_{\alpha\beta}+(\ga^{a} C)_{\alpha\beta}(\ga^b C)_{\ga\delta}\bigg)
\labell{ampc},
\eeqa

with $L_3$  as
\beqa
L_3&=&(2)^{-2(t+s+u)+1}\pi{\frac{\Gamma(-u+\frac{1}{2})
\Gamma(-s+\frac{1}{2})\Gamma(-t+\frac{1}{2})\Gamma(-t-s-u+1)}
{\Gamma(-u-t+1)\Gamma(-t-s+1)\Gamma(-s-u+1)}}
\label{Ls2}
\eeqa

\vskip 0.1in

Note that $L_3$  does not include  any singularities  as the expansion is low energy expansion and all the gamma functions appearing in $L_3$, have no singularities. It is seen that the amplitude involves infinite $s$,$t$,$u$ and $(t+s+u)$ channel poles
and once we are dealing with all massless strings, the expansion is just low energy expansion in which by sending $\alpha'$ to zero we should be able to get to the desired poles in field theory side  ( $t+s+u=-p^ap_a$).

More clearly one of the ideas  for deriving exact form of the amplitude is to be  able to discover all the infinite higher derivative corrections to SYM couplings and also to fix the coefficients of these couplings by producing the S-matrix element. Notice that our amplitude is antisymmetric under interchanging the fermions. Now we try to produce all the infinite poles by exploring new SYM couplings as well as their all order $\alpha'$ higher derivative corrections.

\section{ Infinite $u-$channel singularities and their contact interactions for $p=n$ case }

Unlike the  $<V_{C} V_{\phi} V_{\bar\psi} V_{\psi}>$  , the closed form of the $<V_{C} V_{A} V_{\bar\psi} V_{\psi}>$ correlators just gives us infinite  u-channel massless scalar poles. $st L_1$ expansion is

\beqa
 stL _1&=&-\pi^{3/2}\bigg[\sum_{n=-1}^{\infty}b_n\bigg(\frac{1}{u}(t+s)^{n+1}\bigg)+\sum_{p,n,m=0}^{\infty}e_{p,n,m}u^{p}(st)^{n}(s+t)^m\bigg]
 \labell{highcaap}.
\eeqa
with
\beqa
&&b_{-1}=1,\,b_0=0,\,b_1=\frac{1}{6}\pi^2,\,b_2=2\z(3),e_{0,1,0}=2\z(3),e_{1,0,0}=\frac{1}{6}\pi^2,e_{0,0,1}=\frac{1}{3}\pi^2,
\nonumber
\eeqa
 $b_n$ coefficients do have universal structure \cite{Hatefi:2010ik}. Let us  consider the first term of  ${\cal A}_{22}$ which has infinite $u$-channel poles. One may think $\gamma^\lambda$ in  $\Gamma^{ba\lambda }$ can have both world volume $(\lambda=a)$ and transverse components $(\lambda=i)$ and accordingly there can be both u-channel gauge and scalar pole (sounds ambiguity),  however,  since the integrations for the Chern-Simons action should be taken on the world volume (the sum of the world volume direction should be $p+1$) and
 there is no coupling between one RR, $(p-3)$  form field  $(C_{p-3})$ and one gauge field and two fermions
 (even though

 \beqa
 \frac{(2\pi\alpha')^2\mu_p}{2!(p-2)!}\int d^{p+1}\sigma
 \Tr ( C_{(p-3)}\wedge F\wedge F )\nonumber\eeqa

  is allowed), we conclude that $\lambda$  can not have component in world volume direction. Thus all u-channel scalar poles for $p=n$ case in string theory should be written as

 \beqa
 \frac{\mu_p }{(p)!}
(\veps^v)^{a_0\cdots a_{p-2}ba}H^{i}_{a_0\cdots a_{p-2}}
  2\pi \xi_{1a}(2ik_{1b}) \bu_1^{\alpha}
 (\gamma_i) _{\alpha\beta} u_2^{\beta}\sum_{n=-1}^{\infty}b_n\bigg(\frac{1}{u}(t+s)^{n+1}\bigg)\Tr(\lam_1\lam_2\lam_3) \nonumber\eeqa

where we have normalized the amplitude by a factor of $\frac{\mu_p}{2\pi^{1/2}}$. In the above the trace is replaced as
\beqa
\Tr(P_{-}\fsH_{(n)}M_p \Gamma^{bai })&=&\frac{32 }{2(p)!} (\veps^v)^{a_0\cdots a_{p-2}ba}H^{i}_{a_0\cdots a_{p-2}} \eeqa
The first u-channel pole (for $n=-1$) can be produced by taking its Feynman rule in field theory side as below

\beqa
{\cal A}&=& V_{\alpha}^{i}(C_{p-1},A_1,\phi)G_{\alpha\beta}^{ij}(\phi) V_{\beta}^{j}(\phi,\bPsi_1,\Psi_2),\labell{amp5452}
\eeqa

Let us just mention that to deal with the field theory of a RR and scalar fields, we need to work out either Wess-Zumino (WZ) terms \cite{Myers:1999ps} or  pull-back  or Taylor-expansion (for extensive discussions \cite{Hatefi:2012wj} is suggested).
\vskip.1in

 By applying Taylor expansion as

 \beqa
i\frac{\lambda^2\mu_p}{(p)!}\int d^{p+1}\sigma\Tr\left(\partial_i C_{(p-1)}\wedge F\phi^i \right)\,
 \label{rr}\eeqa
one can obtain the vertex of $V_{\alpha}^{a}(C_{p-1},A_1,\phi)$
where $\lambda=2\pi\alpha'$ so that
\beqa
 V_{\alpha}^{i}(C_{p-1},A_1,\phi)&=&
 i\frac{\lambda^2\mu_p}{(p)!}  H^{i}_{a_0\cdots a_{p-2}} k_{1a_{p-1}} \xi_{1a_{p}}
(\veps^v)^{a_0\cdots a_{p}}\,\Tr\left(\lambda_1\lambda^{\alpha}\right)\,
 \label{rr22}\eeqa

scalar propagator has been derived by taking into account the kinetic term of scalar fields in DBI action $\bigg(-T_p \frac{(2\pi\alpha')^2}{2}\Tr( D_a \phi^i D^a\phi_i)\bigg)$ as follows
\beqa
 G_{\alpha\beta}^{ij}(\phi) &=&\frac{-i\delta_{\alpha\beta}\delta^{ij}}{T_p(2\pi\alpha')^2
k^2}=\frac{-i\delta_{\alpha\beta}\delta^{ij}}{T_p(2\pi\alpha')^2 u}\label{props}\eeqa

In particular $V_{\beta}^{j}(\phi,\bPsi_1,\Psi_2)$ should be derived by considering the fixed kinetic term of fermion fields $\bigg(-T_p (2\pi\alpha')\Tr(-\bPsi\ga^aD_a\Psi)\bigg)$
as well as extracting  the covariant derivative of fermion field such that

 \beqa
 V^{\beta}_{j}(\bPsi_1,\Psi_2,\phi)&=&T_p(2\pi\alpha')\bu_1^A\ga^j_{AB}u_2^B\left(\Tr(\lam_2\lam_3\lam^\beta)-\Tr(\lam_3\lam_2\lam^\beta)\right)
 \label{fver}\eeqa

Having substituted \reef{rr22},\reef{props}
and \reef{fver} in the field theory amplitude of \reef{amp5452}, we are able to precisely produce the first simple massless scalar u-channel pole, however, as it is obvious from the $st L_1$ expansion , our S-matrix does involve infinite u-channel poles. These infinite singularities can be found by postulating an infinite number of  higher derivative corrections to the vertex of $V_{\alpha}^{i}(C_{p-1},A_1,\phi)$ (note that, scalar propagator and all kinetic terms will not receive any corrections \cite{Hatefi:2010ik,Hatefi:2012wj,Hatefi:2012ve} as they have already been fixed in DBI action, see\cite{Hatefi:2013eia}) as below
 \beqa
i\frac{\lambda^2\mu_p}{p!}\int d^{p+1}\sigma
 \,\sum_{n=-1}^{\infty}b_n(\alpha')^{n+1}\Tr\left(\partial_i C_{(p-1)} \wedge
D^{a_0}\cdots D^{a_n}F D_{a_0}\cdots D_{a_n}\phi^i\right)\,
 \label{mm556}\eeqa

Notice that all commutator terms in the definitions of covariant derivatives in \reef{mm556} should be neglected.  Now the infinite extension of the  corrected vertex operator to all orders in $\alpha'$ is given by

 \beqa
 V_{\alpha}^{i}(C_{p-1},A_1,\phi)&=&
 i\frac{\lambda^2\mu_p}{p!} H^{i}_{a_0\cdots a_{p-2}}  k_{1a_{p-1}}\xi_{1a_{p}}
(\veps^v)^{a_0\cdots a_{p}} \sum_{n=-1}^{\infty}b_n (\alpha'k_1.k)^{n+1} \Tr\left(\lambda_1\lambda^{\alpha}\right)\,
 \label{mm66}\eeqa

where
\beqa
 \alpha' k_1.k=t+s, \quad (k_1+k_2+k_3+p)^a=0,\quad p^a (\veps^v)^{a_0\cdots a_{p-1}a}=0 \nonumber\eeqa
 and we have employed the following standard kinetic terms in superstring theory :

\beqa
-T_p (2\pi\alpha')\Tr\left(\frac{(2\pi\alpha')}{2} D_a \phi^i D^a\phi_i-\frac{(2\pi\alpha')}{4}F_{ab}F^{ba}-\bPsi\ga^aD_a\Psi\right)\labell{kin}
\eeqa

  Momentum conservation in world volume direction as well as the constraint for RR are also used.
Now \reef{mm66} is the so called all order $\alpha'$ extension of \reef{rr}. By replacing \reef{mm66} into \reef{amp5452}, keeping fixed scalar propagator and $V_{\beta}^{j}(\phi,\bPsi_1,\Psi_2)$, we can exactly  derive  all infinite u-channel scalar poles of $C A \bar\psi\psi$. Therefore RR $(p-1)$-form field has just induced an infinite number of higher derivative corrections to an on-shell gauge and one off-shell scalar field.
This is a property of closed string RR which proposes all order extensions to all kinds of BPS \cite{Hatefi:2010ik,Hatefi:2013eia,Hatefi:2012rx,Hatefi:2012ve} and non-BPS open strings \cite{Hatefi:2012wj} where  we have called it the universal  property of all order higher derivative corrections.
 Let us end this section by constructing new coupling  and fixing its coefficient  by comparing it with  all contact interactions in $st L_1$ expansion. First consider the following coupling

 \beqa
 \frac{(2\pi\alpha')^2\mu_p}{(p)!}\int d^{p+1}\sigma
 \Tr ( C_{(p-1)}\wedge F  \bar\psi \gamma^i D_i\psi )\label{newc}\eeqa
In order to be able to produce all contact interactions in $st L_1$ expansion (the second terms inside \reef{highcaap}), one can generalize \reef{newc} to all orders in $\alpha'$ as follows:

\beqa
&&\sum_{p,n,m=0}^{\infty}e_{p,n,m} (\alpha')^{2n+m-2}(\frac{\alpha'}{2})^{p}\frac{(2\pi\alpha')^2\mu_p}{\pi(p)!}\int d^{p+1}\sigma
 \Tr ( C_{(p-1)}\wedge D^{a_1}\cdots D^{a_n}D^{a_{n+1}}\cdots D^{a_{2n}}  \nonumber\\&&\times D^{a_1}\cdots D^{a_m} F(D^aD_a)^p D_{a_1}\cdots D_{a_m}(D_{a_1}\cdots D_{a_n}\bar\psi \gamma^i D_{a_{n+1}}\cdots D_{a_{2n}}D_i\psi ))\label{newc22}\eeqa

\vskip.1in

\subsection{An infinite number of scalar poles  for $p+2=n$ case }

In this section, in order to match an infinite number of massless poles of the string theory amplitude of  $C_{p+1} A\bar\Psi\Psi$ with field theory poles, we need to obtain an infinite number of higher derivative corrections to two fermions (with the same chirality), one scalar and one gauge field in type IIB.
\vskip.1in

More importantly we would like to show that the universal conjecture for all order $\alpha'$ corrections \cite{Hatefi:2012rx} holds for the string amplitudes including fermionic strings as well.

The needed terms for these poles are related to the second and the fourth terms of ${\cal A}_1$, thus by extracting the traces, we are able to write down singular terms in string amplitude as

\beqa {\cal A} &=&- \frac{ 2i\alpha'\pi^{-1/2}\mu_p}{(p+1)!} (\veps^v)^{a_0\cdots a_{p-1}} H^{i}_{a_0\cdots a_{p-1}}\bu_1^{\gamma} (\gamma^i )_{\gamma\delta}u_2^{\delta} \nonumber\\&&\times\bigg[-2k_2.\xi_1s+2k_3.\xi_1 t\bigg] L_2
\Tr\left(\lambda_1\lambda_2\lambda_3\right)\,
 \labell{ampc323}
\eeqa

Notice that since  in  $<V_{C} V_{A} V_{\bar\psi}V_{\psi} >$ we do not have external scalar field, we reveal that these two fermions (with the same chirality), one gauge and one scalar couplings must be found just by comparing them with string theory amplitude.

 The $L_2$ 's expansion is
\beqa
 L _2 &=&-\frac{\pi^{5/2}}{2}\left( \sum_{n=0}^{\infty}c_n(s+t+u)^n\right.
\left.+\frac{\sum_{n,m=0}^{\infty}c_{n,m}[s^n t^m +s^m t^n]}{(t+s+u)}\right.\nonumber\\
&&\left.+\sum_{p,n,m=0}^{\infty}f_{p,n,m}(s+t+u)^p[(s+t)^{n}(st)^{m}]\right)
\labell{highcaap12}.
\eeqa

 which clearly shows that it involves an infinite number of  $(t+s+u)$- channel massless scalar poles.
In \cite{Hatefi:2013eia} we have derived the vertex of one off-shell scalar and one $RR-(p+1)$ form field as well as scalar propagator as below
\beqa
G_{\alpha\beta}^{ij}(\phi) &=&\frac{-i\delta_{\alpha\beta}\delta^{ij}}{T_p(2\pi\alpha')^2
k^2}=\frac{-i\delta_{\alpha\beta}\delta^{ij}}{T_p(2\pi\alpha')^2
(t+s+u)},\nonumber\\
V_{\alpha}^{i}(C_{p+1},\phi)&=&i(2\pi\alpha')\mu_p\frac{1}{(p+1)!}(\veps^v)^{a_0\cdots a_{p}}
 H^{i}_{a_0\cdots a_{p}}\Tr(\lambda_{\alpha}).
\labell{Fey}
\eeqa
 The following rule should be  considered to be able to produce an infinite number of scalar $(t+s+u)-$ channel poles
 \beqa
{\cal A}&=&V_{\alpha}^{i}(C_{p+1},\phi)G_{\alpha\beta}^{ij}(\phi)V_{\beta}^{j}(\phi,\bar\Psi,\Psi,A)\label{amp521}
\eeqa

Moreover if we take the  following couplings,

\beqa
\frac{T_p (2\pi\alpha')^3}{4}\bigg[\bPsi \ga^i D_b\Psi D^a\phi_i F_{ab}+\bPsi \ga^i D_b\Psi F_{ab} D^a\phi_i\bigg] \label{zzxc}
 \eeqa

overlook  the commutator terms inside all the covariant derivative terms, take into account $\Tr(\lam_2\lam_3\lam_\beta\lam_1)$  ordering to the first coupling in \reef{zzxc} and also
apply  $\Tr(\lam_2\lam_3\lam_1\lam_\beta)$ ordering to the second coupling in \reef{zzxc} ($\lam_\beta$ holds for Abelian scalar field) then one can easily find the following vertex operator
\beqa
V_{\beta}^{j}(\phi,\bar\Psi,\Psi,A)&=& -i\frac{T_p (2\pi\alpha')^3}{2} \bar u^\gamma (\ga^j)_{\gamma\delta} u^\delta \bigg( -\frac{t}{2}k_3.\xi_1+\frac{s}{2}k_2.\xi_1\bigg)\Tr(\lam_1\lam_2\lam_3\lam_\beta)
\label{eesddem}\eeqa

where momentum conservation is also used.

\vskip.1in

By replacing \reef{eesddem} into \reef{amp521}, keeping the second term of the expansion of $L_2$ for $n=m=0$ (appeared in \reef{highcaap12}) inside \reef{amp521}, and comparing it with string amplitude, we are able to clarify that just the first simple $(t+s+u)$-channel scalar pole  in  \reef{ampc323}  can be precisely produced.
\vskip.2in

But our string amplitude has infinite  $(t+s+u)$-channel scalar poles, so  keeping fixed simple propagator and $V_{\alpha}^{i}(C_{p+1},\phi)$ (there are no corrections to them), one immediately explores that all infinite massless scalar poles should be derived by making use of  all order corrections to two on-shell fermions, one on-shell gauge and  one off-shell scalar field of type IIB as follows:

\beqa
&&{\cal L}^{n,m}= \pi^3\alpha'^{n+m+3}T_p\bigg(a_{n,m}\Tr\bigg[\cD_{nm}\left(\bPsi \ga^i D_b\Psi D^a\phi^i F_{ab} \right)+\cD_{nm}\left( \bPsi \ga^i D_b\Psi F_{ab} D^a\phi^i \right)
\nonumber\\&&+h.c \bigg]
+i b_{n,m}\Tr\bigg[\cD'_{nm}\left(\bPsi \ga^i D_b\Psi D^a\phi^i F_{ab} \right)+\cD'_{nm}\left(
 \bPsi \ga^i D_b\Psi F_{ab} D^a\phi^i  \right)+h.c.\bigg]
\bigg)
 \labell{Lnm681}
\eeqa

where the  following definitions for  higher derivative operators  of $\cD_{nm} ,\cD'_{nm}$  should be taken as well
\beqa
\cD_{nm}(EFGH)&\equiv&D_{b_1}\cdots D_{b_m}D_{a_1}\cdots D_{a_n}E  F D^{a_1}\cdots D^{a_n}GD^{b_1}\cdots D^{b_m}H\nonumber\\
\cD'_{nm}(EFGH)&\equiv&D_{b_1}\cdots D_{b_m}D_{a_1}\cdots D_{a_n}E   D^{a_1}\cdots D^{a_n}F G D^{b_1}\cdots D^{b_m}H\nonumber
\eeqa

First, let us deal with the terms carrying $a_{n,m}$ coefficients. If we consider  the first and the second term of \reef{Lnm681} as well as their hermition conjugate (with the suitable ordering,  mentioned earlier on), we obtain

\beqa
V_{\beta}^{j}(\phi,\bar\Psi,\Psi,A)&=& -i\frac{T_p (2\pi\alpha')^3}{4} \bar u^\gamma (\ga^j)_{\gamma\delta} u^\delta (t^ms^n+t^ns^m)\Tr(\lam_1\lam_2\lam_3\lam_\beta) a_{n,m}
\nonumber\\&&\times
\bigg( -\frac{t}{2}k_3.\xi_1+\frac{s}{2}k_2.\xi_1\bigg)\label{eesddem21}\eeqa

\vskip.1in

Now if we would replace \reef{eesddem21} into \reef{amp521} and would keep the second term of the expansion of $L_2$ for general $n,m$  inside \reef{amp521}, we would be able to show that all infinite $(t+s+u)$-channel scalar poles of string amplitude \reef{ampc323} are exactly produced in field theory as well.

\vskip.2in

If we do the same standard field theory techniques (with the above ordering) to the terms carrying the coefficients of  $b_{n,m}$ then  we derive the following vertex to all orders of $\alpha'$

\beqa
V_{\beta}^{j}(\phi,\bar\Psi,\Psi,A)&=& -i\frac{T_p (2\pi\alpha')^3}{4} \bar u^\gamma (\ga^j)_{\gamma\delta} u^\delta (t^mu^n+s^mu^n)\Tr(\lam_1\lam_2\lam_3\lam_\beta) b_{n,m}
\nonumber\\&&\times
\bigg( -\frac{t}{2}k_3.\xi_1+\frac{s}{2}k_2.\xi_1\bigg)\label{eesddem2}\eeqa

Now by applying on-shell condition
 $(t+s+u=0)$ at each order of $\alpha'$
 to the above vertex, one can easily see that  the common coefficient in both string and field theory amplitudes is precisely re-constructed. It  means that all order $\alpha'$ corrections for the terms including $b_{n,m}$ coefficients are exact.

\vskip.2in

On the other hand , one can show that  all order $\alpha'$ corrections to two fermions and two scalar fields of type IIB are given by

\beqa
&&{\cal L}^{n,m}= \pi^3\alpha'^{n+m+3}T_p\bigg(a_{n,m}\Tr\bigg[\cD_{nm}\left(\bPsi \ga^a D_b\Psi D^a\phi^i D^b\phi_i \right)+\cD_{nm}\left( D^a\phi^i D^b\phi_i  \bPsi \ga^a D_b\Psi \right)
\nonumber\\&&+h.c \bigg]
+i b_{n,m}\Tr\bigg[\cD'_{nm}\left(\bPsi \ga^a D_b\Psi D^a\phi^i D^b\phi_i \right)+\cD'_{nm}\left(
 D^a\phi^i D^b\phi_i  \bPsi \ga^a D_b\Psi \right)+h.c.\bigg]
\bigg)\labell{Lnm}
\eeqa

 \vskip.1in

It is worth trying to  point out some comments on two fermion-two gauge couplings for which they carry three momenta in world volume directions. Consider the following coupling

\beqa
\bPsi \ga^a D_a\Psi F_{bc} F^{bc}\label{UUU}\eeqa
If we take all possible orderings $\Tr(\lambda_2\lambda_3\lambda_1\lambda_{\beta})$ and $\Tr(\lambda_2\lambda_3\lambda_{\beta}\lambda_1)$ where $\lambda_{\beta}$ is related to the Abelian gauge field then we obtain the vertex of two on-shell fermion fields and one on-shell /one off-shell gauge field in field theory side
as

\beqa
v_{b}^{\beta}(\bar\Psi_2,\Psi_3,A_1,A)&=&\bar u \ga^a u (-ik_{3a})   \bigg[(t+s) \xi^{b}  +2k^{b}_1(-k_2.\xi_1-k_3.\xi_1)\bigg] \Tr(\lambda_1\lambda_2\lambda_3\lambda_{\beta})
\label{nml}\eeqa

However, if we apply on-shell equation for the fermion fields to \reef{nml} then we understand that the above coupling \reef{UUU} does not have any contribution to the field theory amplitude.  By the same analysis, one can explicitly show that
$\bPsi \ga^a D_b\Psi F_{bc} F^{ac}$ produces some extra terms which do not appear in the string theory amplitude of $C A\bPsi\Psi$ of type IIB. Therefore we conclude that there are no $\alpha'$ corrections to two fermion-two gauge field couplings of type IIB. One may try to find out this S-matrix in type IIA to see whether or not there are $\alpha'$ corrections to IIA theory \cite{Hatefi:2014lva,Hatefi:2014saa}.

\section{ An infinite number of $t,s$-channel fermion poles }

If we consider the fact that the exchanged strings must have just non-zero fermion number then  we believe neither gauge/scalar nor any other strings (except fermions)  can be propagated. Therefore  to be able to find all infinite $t,s$ channel poles , fermions with the same chirality should be propagated. If we would simplify all the terms appearing in our S-matrix and make use of various identities then  we  could find out that both all  infinite  $t$ and $s$-channel poles make sense just for $p=n$ case and at the end of the day they do come from $A_{26}$ as follows:

\vskip.1in

\beqa
{\cal A} &=& \frac{\alpha'\mu_p \pi\xi_{1a}(2ik_{1b}) }{(p)!} \bu_1^{A} (\ga^b)_{AB}  u_2^{B}  \sum_{n=-1}^{\infty}b_n \frac{1}{t}(u+s)^{n+1}\nonumber\\&&\times
(\veps^v)^{a_0\cdots a_{p-1}a}H_{a_0\cdots a_{p-1}}\Tr(\lam_1\lam_2\lam_3)
 \label{nmm1}\eeqa

Note that  in the above we have also all infinite $s$-channel poles, however, as we know  the amplitude is antisymmetric with respect to  $(s\leftrightarrow t)$. Thus we just produce all t-channel poles.

\vskip.1in

The Feynman rule for producing all infinite fermionic t-channel poles is

\beqa
{\cal A}&=& V_{\alpha}(C_{p-1}, \Psi_3,\bar\Psi)G_{\alpha\beta}(\Psi) V_{\beta}(\Psi,\bar\Psi_2,A_1),\labell{amp560}
\eeqa

To produce the fermionic propagator, one needs to make use of the last term of \reef{kin}. Note that in order to be able to read off $V_{\beta}(\Psi,\bar\Psi_2,A_1)$, one must  extract  the covariant derivative of fermion inside its kinetic term ( $D^a\psi= \partial^a\psi-i[A^a,\psi]$) and in particular take into account  all the desired orderings
of the fermions and the scalars, such that

  \beqa
 V_{\beta}(\Psi,\bar\Psi_2,A_1)&=& -iT_p(2\pi\alpha')\bu_1^A\ga^a_{A}\xi_{1a}\left(\Tr(\lam_1\lam_2\lam^\beta)-\Tr(\lam_2\lam_1\lam^\beta)\right)\nonumber\\
 G_{\alpha\beta}(\psi) &=&\frac{-i\delta_{\alpha\beta}}{T_p(2\pi\alpha')
\fsk}=\frac{-i\delta_{\alpha\beta}\ga^a(k_1+k_2)_a}{T_p(2\pi\alpha') t}
 \label{mm55}\eeqa

\vskip.1in

  Now if  we take the following coupling of  one on-shell RR $(p-1)-$ form field, an on-shell/ an off-shell fermion  then $V_{\alpha}(C_{p-1},\bar \Psi,\Psi)$ can be explored as

    \beqa
i\frac{(2\pi\alpha')\mu_p}{(p)!}
\Tr\left( C_{a_0\cdots a_{p-2}} \bPsi \ga^b \partial_b \Psi \right) (\veps^v)^{a_0\cdots a_{p-2}}\,
\label{nn22}\eeqa

where one has to keep in mind the equations of motion for fermions $(\fsk_{2a}\bu=\fsk_{3a}u=0 )$ as well.
\vskip.1in

 One can now apply some  field theory methods to above coupling \reef{nn22}  to be able to discover the vertex of one RR $(p-1)$-form field and an on-shell/ an off-shell fermion  as

\beqa
 V_{\alpha}(C_{p-1}, \Psi_3,\bar\Psi)&=&i\frac{(2\pi\alpha')\mu_p}{(p)!}
(\veps^v)^{a_0\cdots a_{p-2}} H^{b}_{a_0\cdots a_{p-2}} \ga^b u_2
\Tr\left(\lambda_3\lambda^{\alpha}\right)\,
 \label{nn33}\eeqa

\vskip.1in

If we  substitute  \reef{mm55} and \reef{nn33}  to the Feynman rule in  \reef{amp560} then the field theory amplitude will give rise just the first massless t-channel fermion pole of the string amplitude (for $n=-1$ in \reef{nmm1}).

\vskip.2in

However, as we can see  the string amplitude has infinite $t,s$-channel fermion poles. To be able to obtain all $t-$ channel fermion poles, we need to propose all  the infinite extensions of the higher derivative corrections to  \reef{nn22}.  Indeed the same idea held here as well, namely the kinetic term of fermion fields is already fixed in the effective action so it will not receive any correction, likewise the simple fermion pole has no correction. Thus
we need to work out an  infinite number of higher derivative corrections to the vertex of
$V_{\alpha}(C_{p-1}, \Psi_3,\bar\Psi)$ as

\beqa
i\frac{(2\pi\alpha')\mu_p}{(p)!}
\sum_{n=-1}^{\infty}b_n(\alpha')^{n+1}\Tr\left( C_{a_0\cdots a_{p-2}} D^{a_0}\cdots D^{a_n}\bPsi \ga^b
D_{a_0}\cdots D_{a_n} \partial_b \Psi \right) (\veps^v)^{a_0\cdots a_{p-2}}\,
\label{nn55}\eeqa

\vskip.1in

 By making use of \reef{nn55},  we are able to define all infinite extensions of  $V_{\alpha}(C_{p-1}, \Psi_3,\bar\Psi)$ to all orders in $\alpha' $ as follows:

\beqa
 V_{\alpha}(C_{p-1}, \Psi_3,\bar\Psi)&=&i\frac{(2\pi\alpha')\mu_p}{(p)!}
(\veps^v)^{a_0\cdots a_{p-2}} H^{b}_{a_0\cdots a_{p-2}} \ga^b u_2
\Tr\left(\lambda_3\lambda^{\alpha}\right)\sum_{n=-1}^{\infty}b_n (\alpha'k_3.k)^{n+1} \label{nn66}\eeqa

\vskip.1in

Two remarks are in order. Basically one has to overlook all the connections inside the definitions of the covariant derivatives and the equations of motion should have been applied to the field theory amplitude to be able to obtain the infinite $
t-$ channel fermion poles of the string amplitude.

\vskip.1in

If we replace  \reef{nn66} into \reef{amp560} then we observe that  all  the infinite either $t$- or $s-$ channel fermion poles of \reef{nmm1} are precisely reproduced.

\vskip.1in

Hence we have seen that not only the RR  $(p-1)$-form field proposed all infinite $\alpha'$ corrections  to  two fermions but also it imposed
all order $\alpha'$ corrections to one on-shell gauge and an off-shell scalar field in the world volume of BPS branes in type IIB.

\vskip.1in

 Therefore we conclude that,  this phenomenon seems to be  universal and indeed is useful to determine all the massless  poles of the higher point functions in type IIB superstring theory.

\vskip.1in

Finally it would be nice to observe whether there are  $\alpha'$ higher derivative corrections to four fermions or to two fermion-two gauge field couplings of type IIA \cite{Hatefi:2014lva}, more significantly to see whether or not this universal conjecture on all order $\alpha'$ corrections of type IIB holds for type IIA.

\section{Conclusions}

In this paper we have carried out the conformal field theory calculations and obtained the entire
$<V_{C}V_{\bar\psi}V_{\psi} V_{A}>$ amplitude in IIB superstring theory.  Unlike
$<V_{C}V_{\bar\psi}V_{\psi} V_{\phi}>$ correlator, here we just found an
infinite number of scalar poles for $ p=n$ case. All infinite $t,s-$ channel fermion poles are also discovered.

 \vskip.1in

 We have seen that $V^{\alpha}_i(C_{p+1},\phi)$ and simple poles do not receive  any corrections.
 Thus an infinite   $(t+s+u)-$ channel massless scalar poles help us in exploring all order $\alpha'$ corrections to
 one off-shell scalar , one on-shell gauge field and two on-shell fermion fields of type IIB. In particular the computations of this paper showed us that there are no $\alpha'$ corrections to two fermion-two gauge field couplings of type IIB superstring theory.

\vskip.2in

As we have clarified, inside the  RR vertex operator  winding modes are not included in non compact space. Hence we come to the fact that one should not apply  T-duality to the previous results. For instance in
$<V_{C}V_{\bar\psi}V_{\psi} V_{\phi}>$ amplitude in type IIB , we derived all infinite corrections to two fermion-two scalar couplings while in this paper using direct computations we have shown that there are no corrections
to two gauge-two fermion fields and more importantly there is no even one single u-channel gauge pole either.
 Therefore one must apply direct calculations and should follow some prescriptions to the S-matrices in superstring theory.

\vskip 0.1in

Our computations are done in such a way that all the propagators have been found by conformal field theory formalism and we used the doubling trick, however, RR has two sectors with $(\alpha_n,\tilde{\alpha}_n)$ oscillators. It is not clear to us how to deal with  $\tilde{\alpha}_n$. Just for the completeness, we refer to  \cite{Billo:2006jm} for further information. Basically one has to use analytic   continuation which means that the closed string must be regarded just as a composite state of the open strings.

\vskip.2in

Therefore  background fields in the DBI effective action must be some functions of SYM fields. One  has to consider all background fields as the composite states and eventually all background fields should include  Taylor expansions as it has been discussed in
 \cite{Myers:1999ps}.

\vspace{.3in}  
\section*{Acknowledgements}

I would like to thank J.Polchinski, I. Klebanov, H.Verlinde, E.Witten, N.Arkani-Hamed, J.Maldacena, S.Giddings, W.Siegel, Z.Bern , M.Douglas, J.Schwarz, P.Horava, O.Ganor, J.J.Heckman and C.Vafa for very useful discussions. The last stages of this work are done during my visit to Institute for advanced study in Princeton-NJ, University of California at Berkeley,CA
,University of California Santa Barbara, Kavli institute (KITP), Harvard University, California Institute of Technology, 452-48, Pasadena, CA 91125, USA, at Simons Center for Geometry and Physics, Stony Brook University,Stony Brook, NY 11794, USA and at CERN.  I also thank  R.C.Myers, W.Lerche, I. Antoniadis, N.Lambert, K.S.Narain, F.Quevedo and   L.Alvarez-Gaume for valuable comments. The author thanks CERN for the hospitality.


\end{document}